\documentclass[12pt,journal]{IEEEtran} 
\usepackage{amsfonts}
\usepackage{amssymb}
\usepackage[cmex10]{amsmath}
\usepackage{graphics}
\usepackage[dvips]{graphicx}

\newtheorem{thrm}{\textbf{Theorem}}

\newtheorem{lemma}{\textbf{Lemma}}

\newcommand{\abs}[1]{\left\vert#1\right\vert}
\newcommand{\norm}[1]{\Vert#1\Vert}

\ifCLASSINFOpdf
  
 \else
 \fi
\begin{document}
\onecolumn
\title{A Probably Approximately Correct Answer to Distributed Stochastic Optimization in a Non-stationary Environment\footnote{A part of this paper is submitted to WCNC-2017.}}
\author{\IEEEauthorblockN{B. N. Bharath and Vaishali P\\}
\IEEEauthorblockA{Dept. of ECE, PESIT Bangalore South Campus,\\ Bangalore 560100, INDIA\\}
E-mail: \texttt{bharathbn@pes.edu}, \texttt{vaishali.p.94@gmail.com}
}

\maketitle

\begin{abstract}
This paper considers a distributed stochastic optimization problem where the goal is to minimize the time average of a cost function subject to a set of constraints on the time averages of a related stochastic processes called penalties. We assume that a delayed information about an event in the system is available as a common information at every user, and the state of the system is evolving in an independent and \emph{non-stationary} fashion. We show that an approximate Drift-plus-penalty (DPP) algorithm that we propose achieves a time average cost that is within $\epsilon>0$ of the optimal cost with high probability. Further, we provide a condition on the waiting time for this result to hold. The condition is shown to be a function of the mixing coefficient, the number of samples ($w$) used to compute an estimate of the distribution of the state, and the delay. Unlike the existing work, the method used in the paper can be adapted to prove high probability results when the state is evolving in a non-i.i.d and non-stationary fashion. Under mild conditions, we show that the dependency of the error bound on $w$ is $\exp\{-c w\}$ for $c>0$, which is a significant improvement compared to the exiting work, where $1/\sqrt{w}$ decay is shown.
\end{abstract}
\section{Introduction} \label{sec:intorduction}
A typical stochastic optimization problem involves minimizing the time average of a cost function subject to a set of constraints on the time average penalties \cite{ciftcioglu2013maximizing, neely2013dynamic}. Both cost and penalties depend on the control action and the state of the system. The solution to such a problem is important due to its applicability in various domains such as communications, signal processing, power grids, inventory control for product assembly systems and dynamic wireless networks \cite{ciftcioglu2013maximizing, neely2013dynamic,neely2016distributed, urgaonkar2011optimal, baghaie2010energy, neelyinfocom2015}. An algorithm known as Drift-Plus-Penalty (DPP) (see \cite{neely2010stochastic,neely2008fairness,georgiadis2006resource,neely2006energy, neely2010efficient, neely2010dynamic}) is known to provide a solution for these problems. At each time slot, the DPP method, an extension of the back-pressure algorithm \cite{tassiulas1992stability,tassiulas1993dynamic}, tries to find a control action that minimizes a linear combination of the cost and the drift. The drift is a measure of the deviation of the penalties from the constraints. The DPP algorithm is shown to achieve an approximately optimal solution even when the system evolves in a non-stationary fashion, and is robust to non-ergodic changes in the state \cite{neely2010stochastic}. Further, it is shown to provide universal scheduling guarantees when the state is arbitrary \cite{neely2010stochastic}. 

The DPP algorithm mentioned above assumes that the control action is taken at a centralized unit where the complete state information is available. However, wireless network and crowd sensing applications require a decentralized control action with delayed and heterogenous state information at each node \cite{han2014distributed,neely2010stochastic}. This requires a decentralized version of the DPP algorithm with theoretical guarantees. The author in \cite{neely2016distributed} considers a relaxed version of the above problem. In particular, assuming i.i.d. states with correlated ``common information," the author in \cite{neely2016distributed} proves that a randomized algorithm that chooses one of the $M$ pure strategies, which is obtained as a solution to a Linear Program (\textbf{LP}), is equivalent to solving the original stochastic optimization problem. This equivalence is used to prove that the proposed approximate distributed DPP in \cite{neely2016distributed} is close to being optimal. Several authors use the above result in various contexts such as crowd sensing \cite{han2014distributed}, energy efficient scheduling in MIMO systems \cite{zhang_wcnc2013}, 
to name a few. However, in several practical applications, the states evolve in a dependent and non-stationary fashion \cite{neely2010efficient}. Thus, the following assumptions about the state made in \cite{neely2016distributed} need to be relaxed: (i) independent and (ii) identically distributed. In this paper, we relax the assumption (ii) above, and unlike \cite{neely2016distributed}, we provide a Probably Approximately Correct (PAC) bound on the performance. We would like to emphasize that extending the analysis in \cite{neely2016distributed} to a non-stationary states is non-trivial. Further, the analysis presented in the paper will be very useful to provide theoretical guarantees (PAC bounds) on the distributed DPP for a general model such as (a) states evolving in a dependent non-stationary fashion, and (b) the available data at each node in the network is heterogenous. The only work that is close to ours is \cite{wei2015sample}. However, the authors in \cite{wei2015sample} consider i.i.d. states, and the decision is centralized. Moreover, the method in \cite{wei2015sample} cannot be directly extended to a problem with non-stationary states. This is because the proof in \cite{wei2015sample} requires the control action to be stationary. Now, we highlight the contribution of our work.

\subsection{Main Contribution of the Paper}\label{subsec:main_contributions}
In this paper, we consider a distributed stochastic optimization problem when the states evolve in an independent but \emph{non-stationary} fashion. In particular, we assume that the state is asymptotically stationary, i.e., at each time slot $t \in \mathbb{N}$, the probability measure $\pi_t$ of the state $\omega(t) \in \Omega$ converges to a probability measure $\pi$ as $t \rightarrow \infty$ in the $\mathcal{L}_1$-norm sense. This assumption makes the extension of the method in \cite{neely2016distributed} non-trivial due to the following reasons. When $\pi_t = \pi$ for all $t \in \mathbb{N}$, the author in \cite{neely2016distributed} proves theoretical guarantees by making use of the equivalence between a \textbf{LP} that is a function of $\pi$ and the original stochastic optimization problem. However, when the probabilities are changing, there is no single \textbf{LP} that can be shown to be equivalent to the original problem. Thus, we show that the original problem is equivalent to a ``perturbed" \textbf{LP}, which is a function of the limiting distribution $\pi$ only. Under mild conditions, we prove that the solution to the perturbed \textbf{LP} is approximately equal to the original problem. We use this result to prove theoretical guarantees for an approximate DPP algorithm that we propose in the paper. Moreover, unlike the previous works, we are more interested in providing sample complexity bound rather than just dealing with the averages. The following are the main thesis of our work:

\begin{enumerate}[]
\item For the above model, we show that with high probability, the average cost and penalties obtained by using the proposed approximate DPP are within constants of the optimal solution and the constraints, respectively, provided $t > \text{a threshold}$, and the stochastic process of cost/penalties induced by the algorithm is ``sufficiently" mixing (see Theorem \ref{thm:mainresult1} and Sec. \ref{sec:sims}). These constants capture the degree of non-stationarity (i.e., $\norm{\pi_t - \pi}_1$), the number of samples $w$ used to compute an estimate of the distribution, and the time $t$. 
\item It was shown in  \cite{neely2016distributed,neely_maxweight_unknownenvironment_2012} that the performance gap goes down as $1/\sqrt{w}$. In contrast, under mild conditions, we show that the performance gap goes down exponentially with $w$, provided the complexity of the probability space from which the ``nature" picks $\pi_t$ and $\pi$, measured in terms of the \emph{metric entropy}, is small \cite{van2000applications}. Further, we provide a new condition for the \emph{almost sure} convergence of the time averages of the cost/penalties in terms of the mixing coefficient that measures the dependency. 
\item We show that due to non-stationarity of the states, the performance gap goes down slowly compared to i.i.d. states. This is captured through $\norm{\pi_t - \pi}_1$ and a term that depends on the measure of complexity of  the probability space averaged with respect to $\pi_t$ (see Theorem \ref{thm:mainresult1}). 
\end{enumerate}
The paper is organized as follows. The system model is presented in Sec. \ref{sec:sys_model}. An approximate DPP Algorithm and related theoretical guarantees are provided in Sec. \ref{sec:alg_result}. The interpretation of the main theoretical findings of the paper, and the simulation results are provided in Sec. \ref{sec:sims}. Sec. \ref{sec:conl} concludes the paper.  


\section{System Model} \label{sec:sys_model}
Consider a system consisting of $N$ users making decisions in a distributed fashion at discrete time steps $t\in \{0,1,2,\ldots\}$. Each user $i$ observes a random state $\omega_i(t) \in \Omega_i$, and a ``common information" $X(t) \in \mathcal{X}$ to make a control decision $\alpha_i(t) \in \mathcal{A}_i$, $i=1,2,\ldots, N$. Here, for each user $i$, $\Omega_i$ and $\mathcal{A}_i$ denote the state space and action space, respectively.  Let $\mathbf{\omega}(t) \triangleq \{\omega_1(t),\omega_2(t),\ldots,\omega_N(t)\} \in \Omega$ and $\alpha(t)\triangleq\{\alpha_1(t),\alpha_2(t),\ldots,\alpha_N(t)\} \in \mathcal{A}$, where $\Omega \triangleq \Omega_1 \times \Omega_2 \times \ldots \times \Omega_N$, and $\mathcal{A} \triangleq \mathcal{A}_1 \times \mathcal{A}_2 \times \ldots \times \mathcal{A}_N$. The decision is said to be \emph{distributed} if (see \cite{neely2016distributed})
\begin{itemize}
\item There exists a function $f_i : \Omega_i \times \mathcal{X} \rightarrow \mathcal{A}_i$, such that 
\begin{equation} \label{eq:dist_condition}
\alpha_i(t)\triangleq f_i(\omega_i(t), X(t)),~ i=1,2,\ldots, N,
\end{equation}
where $X(t)$ belongs to the common information set $\mathcal{X}$.
\item The common information $X(t)$ is independent of $\omega(t)$ for every $t \in \mathbb{N}$.
\end{itemize}

At each time slot $t$, the decision $\alpha(t)$ and the state vector $\omega(t)$ result in a cost $p_0(t)\triangleq p_0(\alpha(t),\omega(t))$ and penalties $p_k(t)\triangleq p_k(\alpha(t),\omega(t))$, $k=1,2,\ldots,K$. The central goal of the paper is to analyze an approximate decentralized solution to the following problem when $\omega(t)$, $t\in \mathbb{N}$ is independent but \emph{non-stationary}
\begin{eqnarray}
\mathbf{P_0:}~& \min_{\alpha(\tau) \in \mathcal{A}: \tau \in \mathbb{N}}& \limsup_{t \rightarrow \infty} \frac{1}{t}\sum_{\tau=0}^{t-1} \mathbb{E}p_0(\tau) \nonumber \\
&\text{s. t. }& \hspace{-1cm}\limsup_{t \rightarrow \infty} \frac{1}{t}\sum_{\tau=0}^{t-1} \mathbb{E}p_k(\tau) \leq c_k,~k=1,2,\ldots, K, \nonumber \\
&& \hspace{-0.8cm} \alpha_i(\tau) \text{ satisfies \eqref{eq:dist_condition}, } i=1,2,\ldots,N, \nonumber
\end{eqnarray} 
where the expectation is jointly with respect to the distribution of the state vector $\omega(t)$ and the decision $\alpha(t)$, $t \in \mathbb{N}$. Let $p^{(opt)}$ be the optimal solution to the problem $\mathbf{P_0}$. Note that the first equation in $\mathbf{P_0}$ represents the time average cost while the second and the third equations represent constraints on the penalties and the decisions, respectively. Informally, we are interested in proving a Probably Approximately Correct (PAC) type result of the following form \cite{wei2015sample}
\begin{itemize}
\item For every $\epsilon_k > 0$, with a probability of at least $1 - \delta_k$, $\frac{1}{t} \sum_{\tau = 0}^{t-1} p_k^{(\approx)}(\tau) \leq c_k + \epsilon_k$ provided $t > $ a threshold, where $p_0^{(\approx)}(\tau)$ and $p_k^{(\approx)}(\tau)$, $k=1,2,\ldots,K$ are the cost and penalties, respectively, of an approximate decentralized scheme at $\tau \in \mathbb{N}$. Here $c_0 \triangleq p^{(\textit{opt})}$ is the optimal cost, and $c_k$, $k=1,2,\ldots,K$ are as defined in $\mathbf{P_0}$.
\end{itemize}
First, unlike the model in \cite{neely2016distributed}, it is assumed that the state $\omega(t)$ evolves in an independent but in a \emph{non-stationary} fashion across time $t$. Further, the distribution of $\omega(t)$ denoted $\pi_t(\omega)$, $\omega \in \Omega$ satisfies the following asymptotic stationarity property.

\textbf{Assumption 1:} Assume that there exists a probability measure $\pi(\omega)$ on $\Omega$ such that $\lim_{t \rightarrow \infty} \norm{\pi_t - \pi}_{1} = 0$.

The bounds that we derive will be a function of the complexity of the probability measure space from which the ``nature" chooses $\pi_t(\omega)$. Let us assume that for each $t \in \mathbb{N}$, $\pi_t$ is chosen from a set $\mathcal{P}$. Assuming that $\mathcal{P}$ is a closed set with respect to the $\mathcal{L}_1$\emph{-norm}, we have $\pi \in \mathcal{P}$. A natural way of measuring the complexity is through the covering number of the set $\mathcal{P}$, which is defined as follows.

\textbf{Definition 1: }(see \cite{van2000applications}) A $\delta$-covering of $\mathcal{P}$ is a set $\mathcal{P}_c \triangleq \{\mathcal{P}_1,\mathcal{P}_2,\ldots,\mathcal{P}_M\} \subseteq \mathcal{P}$ such that for all $\pi^{'} \in \mathcal{P}$, there exists a $\mathcal{P}_i \in \mathcal{P}_c$ for some $i=1,2,\ldots,M$ such that $\norm{\pi^{'} - \mathcal{P}_i}_1 < \delta$. The smallest $M$ denoted $M_{\delta}$ is called the covering number of $\mathcal{P}$. Further, $\mathcal{H}(\mathcal{P},\delta) : = \log M_\delta$ is called the \emph{metric entropy}.

{Note that the $\mathcal{L}_1$-norm can be replaced by the following metric: $$d(\pi,\nu)\triangleq\max_{k=0,1,2,\ldots,K} \sum_{\omega \in \Omega} \abs{\pi(\omega) - \nu(\omega)} \abs{p_k(g^{(m)}(\omega), \omega)},$$ which can potentially tighten the bound. However, for the sake of simplicity, we will proceed with the $\mathcal{L}_1$-norm.} Note that in many practical scenarios, the available data is delayed, and a data of size $w$ delayed by $D$ slots will be used for estimation/inference purposes \cite{neely2016distributed, han2014distributed}. 
Since $p_k(t)$, $k=0,1,2,\ldots,K$ depends on $X(t)$ for all $t$, we have that the stochastic process $p_k(t)$ in general is a dependent sequence. The ``degree" of correlation depends on the algorithm used. For $k=0,1,2\ldots,K$ and $s \in \mathbb{N}$, let $\mathbb{P}^{\texttt{ALG},k}_{t, t+s}$ and $\mathbb{P}^{\texttt{ALG},k}_{t}$ denote the joint and marginal distributions of $(p_k(t), p_k(t+s))$ and $p_k(t)$, respectively, induced by the algorithm $\texttt{ALG}$. The following definition captures the correlation.

\textbf{Definition 2:} The $\beta-\texttt{one}$ mixing coefficient of the process $p_k(t)$, $k=0,1,2,\ldots,K$ is given by
$\beta_{\texttt{ALG},k}(s) \triangleq \sup_{t \in \mathbb{N}} \norm{\mathbb{P}^{\texttt{ALG},k}_{t, t+s} - \mathbb{P}^{\texttt{ALG},k}_{t} \otimes \mathbb{P}^{\texttt{ALG},k}_{t + s}}_\text{TV}$,
where $\mathbb{P}^{\texttt{ALG},k}_{t} \otimes \mathbb{P}^{\texttt{ALG},k}_{t + s}$ denotes the product distribution, and $\norm{*}_\text{TV}$ is the total variational norm.

Note that by Pinsker's inequality, we can bound the mixing coefficient in terms of the KL divergence (see \cite{csiszar2011information} and \cite{cover2012elements}). 
Before stating our first result, let us denote the maximum and minimum values of $p_k(t)$, $k=0,1,2,\ldots,K$ by $p_{\text{max},k}$ and $p_{\text{min},k}$, respectively. Further, let $u_{\text{max},k} \triangleq p_{\text{max},k} - p_{\text{min},k}$. In the following section, we propose an Approximate DPP (ADPP) algorithm with the associated theoretical guarantees. The $\beta-\texttt{one}$ coefficient for the ADPP algorithm will be denoted by $\beta_{\texttt{ADPP},k}(s)$.

\section{Algorithm and Main Results} \label{sec:alg_result}
In the following theorem, we provide our first result that is used to prove the ``PAC type" bound for an ADPP algorithm that will be discussed later in this section. 

\begin{thrm} \label{thm:pacfirstresult}
Given an algorithm $\texttt{ALG}$, for any $\epsilon_k > \frac{1}{t}\sum_{\tau=0}^{t-1} \mathbb{E}p_k(\tau) - c_k$, and for constants $u_t \in \mathbb{N}$ and $v_t \in \mathbb{N}$ such that $v_t u_t = t$, we have 
\begin{equation} \label{eq:mcdiarmid_pac1}
\Pr\left\{\frac{1}{t}\sum_{\tau=0}^{t-1} p_k(\tau) - c_k >  \epsilon_k \right\} \leq u_t \exp\left\{\frac{-2 \epsilon_{t,k}^2 v_t^2}{u_{\text{max},k}^2}\right\} + t \beta_{\texttt{ALG},k}(u_t),
\end{equation}
where $\epsilon_{t,k}\triangleq \epsilon_k + c_k - \frac{1}{t}\sum_{\tau=0}^{t-1} \mathbb{E}p_k(\tau)$. Here, $c_0 = p^{(opt)}$, and $c_k$, $k=1,2,\ldots,K$ are the constraint variables in $\mathbf{P_0}$. 
\end{thrm}
\emph{Proof:} See Appendix \ref{app:pacfirstresult}. $\blacksquare$

It is evident from the above result that in order to prove an almost sure convergence, we must have $t \beta_{\texttt{ALG},k}(u_t) \rightarrow 0$ as $t \rightarrow \infty$, i.e., the algorithm $\texttt{ALG}$ should induce a process $p_k(t)$ that is sufficiently mixing. In the following subsection, we prove that the optimal solution to $\mathbf{P_0}$ is close to a \textbf{LP}. 

\subsection{Approximately Optimal LP}
The approximate algorithm that we are going to propose chooses one of the \emph{pure strategy} $$\mathbf{S(\omega)} \triangleq \{\mathbf{s}_1(\omega_1), \mathbf{s}_2(\omega_2), \ldots, \mathbf{s}_N(\omega_N)\}$$ based on the common information $X(t)$,
where $\mathbf{s}_i(\omega_i) \in \mathcal{A}_i$, and $\omega_i \in \Omega_i$, $i=1,2,\ldots,N$. The control action $\alpha_i(t)$ at the user $i$ is chosen as a deterministic function of $\omega(t)$, i.e., $\alpha_i(t)\triangleq \mathbf{s}_i(\omega(t))$ for all $i \in \{1,2,\ldots, N\}$ and for all $t \in \mathbb{N}$. Let the total number of pure strategies be $F\triangleq \prod_{k=1}^N \abs{\mathcal{A}_i}^{\abs{\Omega_i}}$. Enumerating the $F$ strategies, we get $\mathbf{S}^m(\omega)$, $m \in \{1,2,\ldots,F\}$. Note that as pointed out in \cite{neely2016distributed}, it is possible to reduce $F$ by a large amount if the problem has specific structure. For each strategy $m \in \{1,2,\ldots,F\}$, define
\begin{eqnarray}
r_{k,\pi^{'}}^{(m)} &\triangleq& \sum_{\omega \in \Omega} \pi^{'}(\omega) p_k(\mathbf{S}^{m}(\omega), \omega), \label{eq:rkpit}
\end{eqnarray}
where $k = 0,1,2,\ldots,K$ and $\pi^{'} \in \mathcal{P}_c$. Equation \eqref{eq:rkpit} represents the average cost/penalties for the strategy $m$ when the underlaying distribution of $\omega(t)$ is $\pi^{'}$. 
As in \cite{neely2016distributed}, we consider a randomized algorithm where the strategy $m \in \{1,2,\ldots,F\}$ is picked with probability $\theta_m$. Here, $\theta_m$ is a function of the common information $X(t)$. The corresponding average cost/penalty at time $t$ becomes
$\mathbb{E} p^{}_k(t) = \sum_{m=1}^F \theta_m \mathbb{E}_{\lambda} p_k(\mathbf{S}^{(m)}(\omega(t)), \omega(t)) 
= \sum_{m=1}^F \theta_m r_{k,\lambda}^{(m)},\nonumber
$
where $\lambda \in \{\pi_t, \pi, \mathcal{P}_i\}$, $i=1,2,\ldots, M_\delta$. From \textbf{Assumption 1}, we know that $\norm{\pi_t - \pi}_1 \rightarrow 0$, as $t \rightarrow \infty$. With dense covering of the space $\mathcal{P}$, we expect that the limiting distribution is well approximated by $\mathcal{P}_i$ for some $i=1,2,\ldots,M_\delta$ in the covering set. In particular, let $\mathcal{P}_{i^{*}} \triangleq \arg \min_{\mathcal{Q} \in \{\mathcal{P}_1,\ldots,\mathcal{P}_{M_\delta}\}} \norm{\pi - \mathcal{\mathcal{Q}}}_1$, and the corresponding distance be $d_{\pi,\mathcal{P}_{i^{*}}} \triangleq \norm{\pi - \mathcal{P}_{i^{*}}}_1 < \delta$. Consider the following \textbf{LP} denoted $\mathbf{LP_{\mathcal{P}_{i^{*}}}}$:
\begin{eqnarray} \label{eq:LP2}
&\min_{\theta_1,\theta_2,\ldots,\theta_F}& \sum_{m=1}^F \theta_m r_{0,\mathcal{P}_{i^{*}}}^{(m)} \nonumber\\
&\text{subject to}& \sum_{m=1}^F \theta_m r_{k,\mathcal{P}_{i^{*}}}^{(m)} \leq c_k, ~k=1,2,\ldots, K \nonumber\\
&&\sum_{m=1}^F \theta_m = 1. \label{eq:lp}
\end{eqnarray}
Also, we assume that the solution to $\mathbf{LP_{\mathcal{P}_{i^{*}}}}$ exists and the optimal cost is absolutely bounded. Further, define
\begin{equation} \label{eq:gxlp}
G(x) \triangleq \inf\left\{\sum_{m=1}^F \theta_m r_{0,\mathcal{P}_{i^{*}}}^{(m)}: \Theta \in \mathcal{C}_{x,\Theta}\right\},
\end{equation}
where $\Theta \triangleq (\theta_1,\theta_2,\ldots, \theta_F)$, and for any $x \geq 0$, $\mathcal{C}_{x,\Theta} \triangleq \{\Theta: \sum_{m=1}^F \theta_m r_{k,\mathcal{P}_{i^{*}}}^{(m)} \leq c_k + x, ~k=1,2,\ldots, K, \text{ and } \Theta \mathbf{1}^T = 1\}$. Note that $G(0)$ corresponds to $\mathbf{LP_{\mathcal{P}_{i^{*}}}}$. We make the following important smoothness assumption about the function $G(x)$ \textbf{LP}. 

\textbf{Assumption 2:} The function $G(x)$ is \emph{Lipschitz continuous} around the origin, i.e., for some $c > 0$, we have
\begin{equation}
\abs{G(x) - G(y)} \leq c \abs{x-y}, \text{ for all } x,y \geq 0.
\end{equation}

In the theorem to follow, given that \textbf{Assumption 2} is valid, we prove that the optimal cost of the linear optimization problem in \eqref{eq:LP2} is ``close" to the optimal cost of $\mathbf{P}_0$. The above assumption can be relaxed at the expense of looser bound on the guarantees. 

\begin{thrm} \label{thm:lppit_popt_relation}
Let $p^{\text{(opt)}}$ and $p^{(\text{opt})}_{\mathcal{P}_{i^{*}}}$ be the optimal solution to the problems $\mathbf{P0}$ and $\mathbf{{LP}_{\mathcal{P}_{i^{*}}}}$, respectively. Then, under \textbf{Assumption 2}, we have $p^{(\text{opt})}_{\mathcal{P}_{i^{*}}} < p^{\text{(opt)}} + (c + 1) \Delta_{\pi,\mathcal{P}_{i^{*}}}$, where $\Delta_{\pi,\mathcal{P}_{i^{*}}} = b_{\text{max},k}  (d_{\pi,\mathcal{P}_{i^{*}}} + \nu)$, and $b_{\text{max},k} \triangleq \max\{\abs{p_{\text{max},k}}, \abs{p_{\text{min},k}}\}$. 
\end{thrm}
\emph{Proof:} See Appendix \ref{app:lppit_popt_relation}. $\blacksquare$

We use the above result to prove guarantees related to an approximate algorithm that will be discussed in the following section.

\subsection{Approximate DPP Algorithm} \label{subsec:dpp}
In this subsection, we present an online distributed algorithm that approximately solves the problem $\mathbf{P0}$. We assume that at time $t \in \mathbb{N}$, all nodes receive feedback specifying the values of all the penalties and the states, namely, $p_1(t-D), p_2(t-D), \ldots, p_K(t-D)$ and $\omega(t-D)$. Recall that $D > 0$ is the delay in the feedback. The following set of queues are constructed using this information
\begin{equation} \label{eq:queue_update}
Q_k(t+1) = \max\{Q_k(t) + p_k(t-D) - c_k, 0\},
\end{equation}
$k=1,2,\ldots,K$, and $t \in \mathbb{N}$. These queues act as the common information, i.e., $X(t) = \mathbf{Q(t)}$, where $\mathbf{Q(t)} \triangleq (Q_1(t),Q_2(t),\ldots,Q_K(t))$. Further, $\omega(t-i)$, $i=D,D+1,\ldots,D+w-1$, will be used to find an estimate of the state probabilities that is required for the algorithm that we propose. For all $k=1,2,\ldots, K$, we let $p_k(t) = 0$, $t = -1,-2,\ldots, -D\}$. The \emph{Lyapunov} function is defined as 
\begin{equation}
\mathcal{L}(t) \triangleq \frac{1}{2} \norm{\mathbf{Q}(t)}^2_2 = \frac{1}{2} \sum_{i=1}^K Q_i^2(t),
\end{equation}
and the corresponding drift is given by $\Delta(t)\triangleq\mathcal{L}(t+1) - \mathcal{L}(t)$ for all $t \in \mathbb{N}$. The following lemma provides an upper bound on the DPP: $\mathbb{E} \left[\Delta(t+D) + V p_0(t) \left \vert \right. \mathbf{Q}(t)\right]$, $V \geq 0$ that will be used in the algorithm to follow. The proof of the lemma follows  directly from the proof of Lemma $5$ of \cite{neely2016distributed}, and hence omitted.  

\begin{lemma} 
For a fixed constant $V \geq 0$, we have
\begin{eqnarray} \label{eq:dpp_bound_lemma}
&&\hspace{-1.0cm}\mathbb{E} \left[\Delta(t+D) + V p_0(t) \left \vert \right. \mathbf{Q}(t)\right] \leq B_t(1 + 2D) \nonumber \\ && \hspace{-1.4cm}+ V \sum_{m=1}^F \beta_m(t) r_{0,\pi_t}^{(m)} + \sum_{k=1}^K Q_k(t) \left[ \sum_{m=1}^F \beta_m(t) r_{k,\pi_t}^{(m)} - c_k\right],
\end{eqnarray}
where $r_{k,\pi_t}^{(m)} \triangleq \sum_{\omega \in \Omega} \pi_t(\omega) p_k(\mathbf{S}^{(m)}(\omega), \omega)$, $k=0,1,2,\ldots,K$, and 
\begin{equation} \label{eq:btau}
~B_t\triangleq \max_{m \in \{1,2,\ldots, F\}} \frac{1}{2} \sum_{k=1}^K \sum_{\omega \in \Omega} \pi_t(\omega) \abs{p_k(\mathbf{S}^{(m)}(\omega), \omega) - c_k}.
\end{equation}
\end{lemma}

Note that as $t \rightarrow \infty$, $B_t \rightarrow B$. The expression for $B$ can be obtained by replacing $\pi_t(\omega)$ by $\pi(\omega)$ in the expression for $B_t$. In the following, we provide the algorithm. 
\begin{itemize}
\item \textbf{Algorithm:} Given the delayed feedback of size $w$, i.e., $Q_k(t-i-D)$, $i=0,1,\ldots,w-1$ at each time slot $t \in \mathbb{N}$ and  $k=1,2,\ldots,K$, perform the following steps
\begin{itemize}
\item \textbf{Step 1:} Find the probability measure from the covering set $\mathcal{P}_c$ that best fits the data, i.e., pick $\mathcal{P}_{j^*} \in \mathcal{P}_c$ such that 
\begin{equation} \label{eq:detect_covering_alg}
j^* : = \arg \max_{j \in \{1,2,\ldots,M_\delta\}} \frac{1}{w}\sum_{\tau=t-D-w+1}^{t-D} \log \left({\mathcal{P}_{j}(\omega(\tau))}\right).
\end{equation}
\item \textbf{Step 2:} Choose $m \in \{1,2,\ldots, F\}$ (breaking ties arbitrarily) that minimizes the following:
\begin{equation}
V r_{0,\mathcal{P}_{j^*}}^{(m)} + \sum_{k=1}^K Q_k(t) r_{k,\mathcal{P}_{j^*}}^{(m)}. 
\end{equation}
\end{itemize}
\end{itemize}
We say that there is an error in the outcome of step $1$ of the algorithm if $\mathcal{P}_{j^*} \neq \mathcal{P}_{i^*}$. Recall that $i^*$ corresponds to the index of the probability measure in the covering set that is close to $\pi$ in the $\mathcal{L}_1$ norm sense. The  error event $\mathcal{E}_{\delta, t}$, $t \in \mathbb{N}$ is defined as those outcomes for which $j^* \neq i^*$. We make the following assumption which will come handy in the proof of Theorem \ref{thm:mainresult1} below.

\textbf{Assumption 3:} Assume that for all $j=1,2,\ldots,M_\delta$, when $\mathcal{P}_{j}(\omega) \neq 0$, there exist constants $\alpha_\delta >\beta_\delta > 0$, such that $\alpha_\delta > \mathcal{P}_{j}(\omega) > \beta_\delta > 0$ for all $\omega \in \Omega$.

The following theorem uses \eqref{eq:mcdiarmid_pac1} to provide a PAC result for the above algorithm. 
 \begin{thrm} \label{thm:mainresult1}
Under \textbf{Assumptions 1-3}, for the proposed \textbf{Algorithm}, for some finite positive constants $V$, $C$ and $c$, the following holds.
\begin{itemize}
\item \textbf{Part A:} For every $\epsilon > 0$, with a probability of at least $1-\gamma_0$, 
\begin{equation} \label{eq:obj_mainres}
\frac{1}{t}\sum_{\tau=0}^{t-1} p_0(\tau) \leq  p^{(opt)} + (c + 1) \Delta_{\pi,\mathcal{P}_{i^{*}}} + \psi_t(\delta) + \epsilon
\end{equation}
provided $t \in \mathcal{T}_{t,0}$. Here, $\gamma_0 > t \beta_{\texttt{ADPP},0}$.

\item \textbf{Part B: } For every $\epsilon > 0$, with a probability of at least $1-\gamma_1$, 
\begin{equation} \label{eq:constraint_mainres}
\frac{1}{t}\sum_{\tau=0}^{t-1} p_k(\tau) \leq  c_k + Q_{\texttt{up}}(t) + \epsilon, ~k = 1,2,\ldots,K
\end{equation}
provided $t \in \mathcal{T}_{t,1}$. Here $\gamma_1 > t (\max_{k=1,2,\ldots,K}\beta_{\texttt{ADPP},k})$. 
\end{itemize}
In the above, $\mathcal{T}_{t,i} \triangleq \left\{t: t > \frac{u_{\text{max},0} u_t}{\sqrt{2} \epsilon } \sqrt{\log\left(\frac{u_t}{\gamma_i - t \beta_{\texttt{ADPP},i} (u_t)}\right)}\right\}$, $i \in \{0,1\}$, $\Delta_{\pi,\mathcal{P}_{i^{*}}} = b_{\text{max},k}  (d_{\pi,\mathcal{P}_{i^{*}}} + \nu)$, and 
$\psi_t(\delta) \triangleq \frac{V(c+1){\bar{J}}_t + \bar{H}_t + C/t}{V}+  \frac{1+2D}{tV}\sum_{\tau=0}^{t-1} B_\tau P_{e,\texttt{up}}^{(\tau)} + \frac{p_{\text{max},0}}{t} \sum_{\tau=0}^{t-1} P_{e,\texttt{up}}^{(\tau)} $,
where $$\bar{J}_t \triangleq \max_{0\leq k \leq K} p_{\text{max},k} \left(\frac{1}{t}\sum_{\tau = 0}^{t-1}\norm{\pi_\tau - \pi}_1 + \delta \right)$$ $\bar{H}_t \triangleq \frac{1+2D}{t}\sum_{\tau = 0}^{t-1} B_\tau$, $\mathcal{D}_{\tau,j}  \triangleq \frac{1}{w} \sum_{s=\tau-D}^{\tau-D-w} \mathbb{E}_{\pi_\tau} \log \left(\frac{\mathcal{P}_{j}(\omega(s))}{\mathcal{P}_{i^*}(\omega(s))}\right)$, $\mathcal{D}_\tau \triangleq \min_{j \neq i^*}\mathcal{D}_{\tau,j}$, $\zeta_\delta \triangleq {\left[\log\left(\frac{\alpha_\delta}{\beta_\delta}\right)\right]^2}$, and $P_{e,\texttt{up}}^{(\tau)}\triangleq\exp\left\{- {2 \zeta_\delta \mathcal{D}_\tau^2 w} + \mathcal{H}(\mathcal{P},\delta) \right\}$, $0 < \alpha_\delta < \beta_\delta$.
Further, 
$Q_{\texttt{up}}(t)\triangleq \sqrt{\frac{V F}{t} + \frac{\Gamma_t}{ t^2}},$
$$\Gamma_t \triangleq V(c + 1) (\Delta_{\pi,\mathcal{P}_{i^{*}}} + {{\bar{J}_t}) + \bar{H}_t + C} + (1 + 2D) \sum_{\tau=0}^{t-1} B_\tau P_{e,\texttt{up}}^{(\tau)} + p_{\text{max},0} \sum_{\tau=0}^{t-1} P_{e,\texttt{up}}^{(\tau)}$$ and $p_{\text{max,k}}$, $k=1,2,\ldots,K$ is as defined earlier.
\end{thrm}
\emph{Proof:} See Appendix \ref{app:mainresult1}. $\blacksquare$

\section{Interpretation and Simulation Results} \label{sec:sims}
The following observations are made:
\begin{itemize}
\item As in \cite{neely2016distributed}, \textbf{Part A} and \textbf{Part B} of Theorem \ref{thm:mainresult1} show the tradeoff between satisfying the constraints and minimizing the objective captured by the constant $V\geq 0$. Note that $\gamma_0$ and $\gamma_1$ are lower bounded by $t$ times the $\beta$-\texttt{one} mixing coefficient. If the algorithm induces sufficient mixing, i.e., $\beta_{\texttt{ADPP},k}(t) \propto 1/t^\alpha$, where $\alpha > 1$, then asymptotically, the lower bound on $\gamma_0$ and $\gamma_1$ go to zero. This leads to a new condition for \emph{almost sure} convergence of the time average cost/penalties, i.e., if $t \beta_{\texttt{ADPP},k} \rightarrow 0$ for all $k=0,1,2,\ldots, K$, then the result in \eqref{eq:obj_mainres} and \eqref{eq:constraint_mainres} hold \emph{almost surely} \cite{wei2015sample} with the expense of increased waiting time (see the definition of $\mathcal{T}_{t,i}$ above). Unlike the results in \cite{wei2015sample}, the method here can be applied for non-stationary dependent state process and the condition for the almost sure convergence is in terms of the mixing coefficient. 
\item
When $\omega(t)$ is i.i.d., both $\psi_\delta(t)$ and $Q_{\texttt{up}}(t)$ reduces, leading to a smaller objective value and a better constraints satisfaction capability. This is due to the fact that $\norm{\pi_\tau - \pi} = 0$ for all $\tau$ which reduces the value of $\bar{J}_t$. 
\item 
Note that unlike \cite{neely_maxweight_unknownenvironment_2012}, the dependency on $w$ is exponential instead of $\frac{1}{\sqrt{w}}$. Also, higher metric entropy, $\mathcal{H}(\delta,\mathcal{P})$ requires larger values of $w$ for better performance. Equivalently, when the complexity of the model, $\mathcal{P}$ is low, then the learnability improves. Thus, as $t \rightarrow \infty$, we have a better result compared to \cite{neely2016distributed,neely_maxweight_unknownenvironment_2012}.
\item 
As $t \rightarrow \infty$, $\bar{H}_t$ goes to $B(1+2D)$ and $\bar{J}_t$ goes to a constant. Further, both terms $\frac{1+2D}{tV}\sum_{\tau=0}^{t-1} B_\tau P_{e,\texttt{up}}^{(\tau)}$ and $\frac{p_{\text{max},0}}{t} \sum_{\tau=0}^{t-1} P_{e,\texttt{up}}^{(\tau)}$ go to zero since $\mathcal{D}_\tau $ goes to a constant for a large values of $\tau$. Putting these together, we get the following result. When $t \beta_{\texttt{ADPP},k} \rightarrow 0$ for all $k=0,1,2,\ldots, K$,  
$\limsup_{t \rightarrow \infty} \frac{1}{t}\sum_{\tau=0}^{t-1} p_0(\tau) \leq  p^{(opt)} + (c + 1) \Delta_{\pi,\mathcal{P}_{i^{*}}} + \text{constant} + \epsilon$ and $\limsup_{t \rightarrow \infty}\frac{1}{t}\sum_{\tau=0}^{t-1} p_k(\tau) \leq  c_k + \text{constant} + \epsilon, ~k = 1,2,\ldots,K$ hold with probability one. This shows that the error will not be zero even when $t \rightarrow \infty$. 
\end{itemize}

\subsection{Simulation Results}
\begin{figure}[ht]
    \centering
    \includegraphics[width=14.0cm,height =10.0cm]{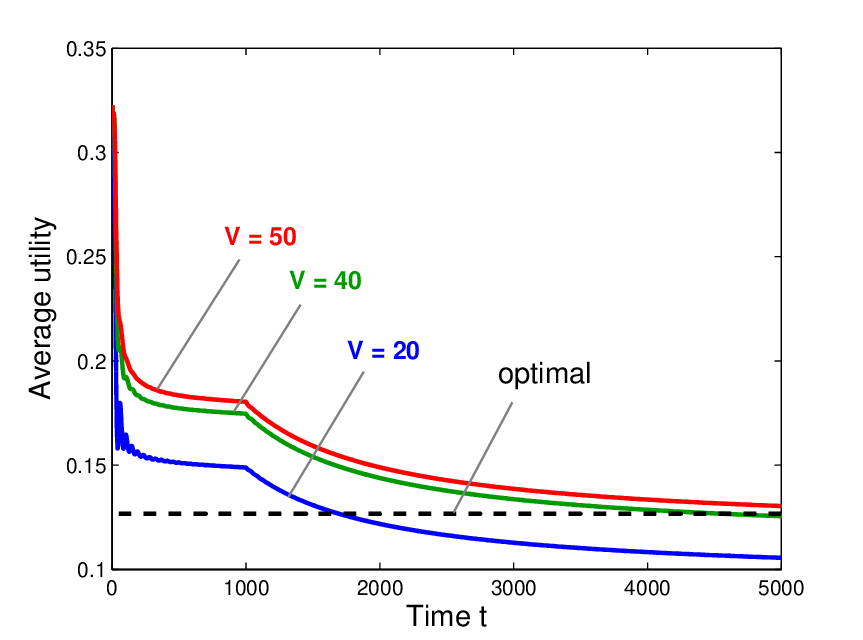}
    \caption{Figure shows the plot of the utility versus time.}
    \label{fig:utility_vs_time}
\end{figure}
\begin{figure}[ht]
    \centering
    \includegraphics[width=14.0cm,height = 10.0cm]{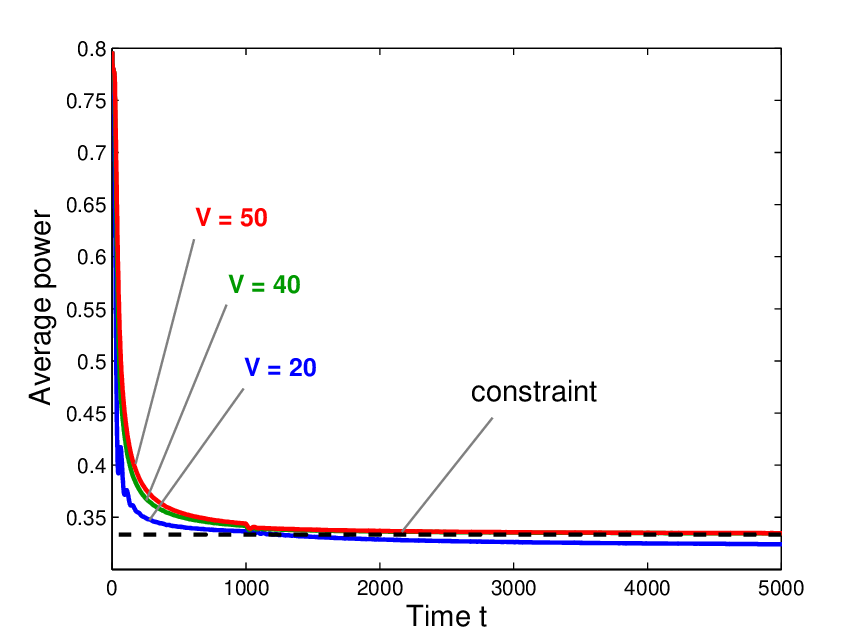}
    \caption{Figure shows the plot of the average power versus time.}
    \label{fig:power_vs_time}
\end{figure}

For the simulation setup, we consider a system comprising of $3$ sensors observing the state $$\omega(t)\triangleq\{\omega_1(t),\omega_2(t), \omega_3(t)\} \in \{0,1,2,3\}^3,$$ and reports the observation to the central unit, and the reporting incurs a penalty. The problem is to maximize the utility given by $$u_0(t)\triangleq \min\left\{\frac{\alpha_1(t) \omega_1(t)}{10} + \frac{\alpha_2(t) \omega_2(t) + \alpha_3(t) \omega_3(t)}{20},1\right\},$$ where $\alpha_i(t) \in \{0,1\}$, $i=1,2,3$ is the decision variable. {Here, the utility is the negative of $p_0(t)$.} The probability measure $\pi_t$ is chosen from a set of $8$ distributions, and converges to $ \Pr(\omega_i=0) = 0.1$, $\Pr(\omega_i=1) = 0.7$, $\Pr(\omega_i=2) = 0.1$, $\Pr(\omega_i=3) = 0.1$, $i=1,2,3$. The details of the distribution that is used in the transient time will be provided in the full length version of this paper. The optimal value of this is $p^{(opt)} = 0.1267$. We have run the simulation for $5000$ time slots and averaged over $2000$ instantiations. When $\alpha_i(t) = 1$, $i=1,2,3$, a power of $1$ watt each is consumed. We assume an average power constraints of $1/3$ at each node. Figures \ref{fig:utility_vs_time} and \ref{fig:power_vs_time} show the plots of utility and penalty versus time $t$ for different values of $V$, $D=10$ and $w=40$, demonstrating the tradeoff in terms of $V$.  For large values of $t$, the utility achieved by the algorithm with $V=50$ is close to optimum while satisfying the constraints thereby confirming the optimality of the algorithm. It is important to note that the mixing coefficient can be easily estimated, and hence mixing condition can be verified through the simulation. A mathematically rigorous analysis of verifying the mixing condition will be a part of our future work. 

\section{Concluding Remarks}\label{sec:conl}
In this paper, we considered the problem of distributed stochastic optimization problem with non-stationary states. Assuming asymptotic stationarity of the states, we showed that the stochastic optimization problem is approximately equal to a linear program that is a function of the limiting distribution. An approximate Drift-Plus-Penalty (DPP) algorithm is proposed to solve the problem. For the proposed algorithm, we showed that with certain probabilities $\gamma_0$ and $\gamma_1$, the average cost and penalties obtained by using the proposed approximate DPP are within constants of the optimal solution and the constraints, respectively, provided $t > \text{a threshold}$. The threshold is in terms of the mixing coefficient that indicates the non-stationarity of the cost/penalties. The approximation errors capture the degree of non-stationarity (i.e., $\norm{\pi_t - \pi}_1$), the number of samples $w$ that is used to compute an estimate of the distribution, and the time $t$. Further, we showed that the error goes down exponentially with $w$, which is a significant improvement compared to the existing work. 
\appendices

\section{Proof of Theorem \ref{thm:pacfirstresult}} \label{app:pacfirstresult}
Let us fix constants $u_t$ and $v_t$ as defined in the theorem. Consider the following set of sequences of the cost/penalties $\mathcal{S}_{i,k}^{(t)} \triangleq \{p_k(j(u_t - 1) +i): j = 0,1,2,\ldots, v_t -1\}$, $i=0,1,2,\ldots,u_t-1$ and $k=0,1,\ldots,K$. Now, the time average of cost/penalties can be written as $\frac{1}{t}\sum_{\tau = 0}^{t-1} p_k(\tau) = \frac{v_t}{t}\sum_{i = 0}^{u_t-1} \Psi_{k,i,t}$, where $\Psi_{k,i,t}\triangleq \frac{1}{v_t}\sum_{s \in \mathcal{S}_{i,k}^{(t)}} p_k(s)$. Note that each term in $\Psi_{i,k,t}$ is at least $u_t$ slots apart. Using this, and $\bar{p}_k(t) \triangleq \frac{1}{t}\sum_{\tau = 0}^{t-1} p_k(\tau)$, the left hand side of \eqref{eq:mcdiarmid_pac1} can be written as
\begin{eqnarray}
\Pr\left\{\bar{p}_k(t) - \mathbb{E}{\bar{p}(t)} > \epsilon_{t,k}\right\} &=& \Pr\left\{\frac{v_t}{t}\sum_{i = 0}^{u_t-1} \Delta \Psi_{k,i,t} > \epsilon_{t,k}\right\} \nonumber\\
&\stackrel{(a)}{\leq}& \sum_{i=0}^{u_t - 1} \Pr\left\{\Delta \Psi_{k,i,t} > \epsilon_{t,k}\right\} \nonumber\\ &\stackrel{(b)}{\leq}& \sum_{i=0}^{u_t - 1} \Pr\left\{\Delta \tilde{\Psi}_{k,i,t} > \epsilon_{t,k}\right\} + t \beta_{\texttt{ALG},k}(u_t), \label{eq:firstbound}
\end{eqnarray}
where $\epsilon_{t,k} \triangleq \epsilon_k + c_k - \mathbb{E}{\bar{p}_k{(t)}}$, $\Delta \Psi_{k,i,t} \triangleq \Psi_{k,i,t} - \mathbb{E}{\Psi_{k,i,t}}$, $\tilde{\Psi}_{k,i,t} \triangleq \frac{1}{v_t}\sum_{\tau \in \mathcal{S}_{i,k}^{(t)}} \tilde{p}_k(\tau)$, and $\tilde{p}_k(\tau)$ is an independent stochastic process having the same distribution as $p_k(\tau)$, $k=0,1,2,\ldots,K$. In the above, $(a)$ follows from the fact that the convex combination of terms being greater than a constant implies that at least one of the term should be greater than the constant, and using the union bound. The inequality $(b)$ is obtained by applying proposition $1$ of \cite{kuznetsov2014generalization} to the indicator function $g \triangleq \mathbf{1}\{\Delta \Psi_{k,i,t}^{(\tau)} > \epsilon_{t,k}\}$. Note that $\Delta \Psi_{k,i,t}^{(\tau)}$ is the sum of $v_t$ independent random variables. Thus, by applying the Mcdiarmids inequality along with the fact that $\Psi_{k,i,t}^{(\tau)} \leq u_{\max,k}$ for all $t \in \mathbb{N}$, we get $$\Pr\left\{\Delta \Psi_{k,i,t}^{(\tau)} > \epsilon_{t,k}\right\} \leq \exp\left\{\frac{-2 \epsilon_{t,k}^2 v_t^2}{u_{\text{max},k}^2}\right\}. $$ Using this in \eqref{eq:firstbound}, we get the desired result. $\blacksquare$

\section{Proof of Theorem \ref{thm:lppit_popt_relation}} \label{app:lppit_popt_relation}
Consider the cost/penalties of the problem $\mathbf{P_0}$
\begin{eqnarray}
\limsup_{t \rightarrow \infty} \frac{1}{t}\sum_{\tau=0}^{t-1} \mathbb{E}p_k(\tau)
&\stackrel{(a)}{=}& \limsup_{{t \rightarrow \infty,} {t > t^{'}}} \left[\frac{1}{t}\sum_{\tau=0}^{t-1} \sum_{\omega \in \Omega} \mathcal{P}_{i^{*}}(\omega) \Phi_k(\tau) + \frac{1}{t}\sum_{\tau=0}^{t-1} \sum_{\omega \in \Omega} \left(\pi_\tau(\omega) -\pi(\omega) \right) \Phi_k(\tau)\right. \nonumber\\
&&\left.+ \frac{1}{t}\sum_{\tau=0}^{t-1} \sum_{\omega \in \Omega} \left(\pi(\omega) -\mathcal{P}_{i^{*}}(\omega) \right) \Phi_k(\tau)\right] 
\end{eqnarray}
for $k = 0,1,2,\ldots,K$ and some $t^{'} > 0$.  In the above, $\Phi_k(\tau) \triangleq p_k(\alpha(\tau), \omega(\tau))$ and $(a)$ follows by adding and subtracting $\mathcal{P}_{i^{*}}$ and $\pi$. Since $\lim_{t \rightarrow \infty} \norm{\pi_t - \pi}_1 = 0$, for every $\nu > 0$, there exists a $t^{'} \in \mathbb{N}$ such that for all $t > t^{'}$, $\norm{\pi_t - \pi}_1 < \nu$. Using the fact that $$\abs{\sum_{\omega \in \Omega} \left(\pi_\tau(\omega) -\pi(\omega) \right) \Phi_k(t)} \leq \sum_{\omega \in \Omega} \abs{\left(\pi_t(\omega) -\pi(\omega) \right)} \abs{\Phi_k(t)} \leq \max\{\abs{p_{\text{max},k}}, \abs{p_{\text{min},k}}\}  \nu$$ for every $t > t^{'}$, we have
\begin{equation} \label{eq:util_penal_term2}
-b_{\text{max},k} \nu \leq \frac{1}{t}\sum_{\tau=0}^{t-1}\sum_{\omega \in \Omega} \left(\pi_\tau(\omega) -\pi(\omega) \right) \Phi_k(t) \leq b_{\text{max},k} \nu,
\end{equation}
where $b_{\text{max},k} \triangleq \max\{\abs{p_{\text{max},k}}, \abs{p_{\text{min},k}}\}$. Similarly, we have 
\begin{equation} \label{eq:util_penal_term3}
- b_{\text{max},k} d_{\pi,\mathcal{P}_{i^{*}}} \leq \frac{1}{t}\sum_{\tau=0}^{t-1} \sum_{\omega \in \Omega} \left(\pi(\omega) -\mathcal{P}_{i^{*}}(\omega) \right) \Phi_k(\tau) \leq b_{\text{max},k} d_{\pi,\mathcal{P}_{i^{*}}}.
\end{equation}
Using the above bounds, we get the following lower bound for all $k=1,2,\ldots,K$.
\begin{equation}
\limsup_{t \rightarrow \infty} \frac{1}{t}\sum_{\tau=0}^{t-1} \mathbb{E}p_k(\tau) \geq \limsup_{t \rightarrow \infty} \frac{1}{t}\sum_{\tau=0}^{t-1} \sum_{\omega \in \Omega} \mathcal{P}_{i^{*}}(\omega)\Phi_k(\tau) - \Delta_{\pi,\mathcal{P}_{i^{*}}},
\end{equation}
where $\Delta_{\pi,\mathcal{P}_{i^{*}}} = b_{\text{max},k}  (d_{\pi,\mathcal{P}_{i^{*}}} + \nu)$. By using the above lower bound in $\mathbf{P_0}$, we get the following optimization problem denoted $\mathbf{P_1}$ 
\begin{eqnarray}
&\min_{\alpha(\tau) \in \mathcal{A}: \tau \in \mathcal{N}}& \limsup_{t \rightarrow \infty} \frac{1}{t}\sum_{\tau=0}^{t-1} \mathbb{E} p_0(t) - \Delta_{\pi,\mathcal{P}_{i^{*}}} \nonumber \\
&\hspace{-1.22cm}\text{s. t. }& \hspace{-1.6cm}\limsup_{t \rightarrow \infty} \frac{1}{t}\sum_{\tau=0}^{t-1} \mathbb{E}p_k(t) \leq c_k + \Delta_{\pi,\mathcal{P}_{i^{*}}},~k=1,2,\ldots, K, \nonumber \\
&& \hspace{-0.8cm} \alpha_i(t) \text{ satisfies \eqref{eq:dist_condition}, } i=1,2,\ldots,K, \nonumber
\end{eqnarray} 
where the expectation is taken with respect to $\mathcal{P}_{i^{*}}$. Note that the optimal cost obtained by solving $\mathbf{P_1}$ is smaller than $p^{\text{opt}}$. Further, the term $\Delta_{\pi,\mathcal{P}_{i^{*}}}$ is independent of the control action. It is evident from $\mathbf{P_1}$ that it is equivalent to $\mathbf{P_0}$ where the states $\omega(t)$ is i.i.d. whose distribution is $\mathcal{P}_{i^{*}}$. Using Theorem $1$ of \cite{neely2016distributed}, it can be shown that the solution to $\mathbf{P_1}$ is equal to $G(\Delta_{\pi,\mathcal{P}_{i^{*}}}) - \Delta_{\pi,\mathcal{P}_{i^{*}}}$,where $G(x)$ is as defined in \eqref{eq:gxlp}. 
Note that when $\Delta_{\pi,\mathcal{P}_{i^{*}}} =0$, we get back the original problem in $\mathbf{LP_{\mathcal{P}_{i^{*}}}}$. Thus, from \textbf{Assumption 2}, we have that $$\abs{p^{\text{(pert)}}_{\mathcal{P}_{i^{*}}} - p^{(\text{opt})}_{\mathcal{P}_{i^{*}}}} < c \Delta_{\pi,\mathcal{P}_{i^{*}}} + \Delta_{\pi,\mathcal{P}_{i^{*}}}= (c+1) \Delta_{\pi,\mathcal{P}_{i^{*}}},$$ where $p^{\text{(pert)}}_{\mathcal{P}_{i^{*}}}$ denotes the optimal cost of the above problem. This leads to the following bound $p^{\text{(pert)}}_{\mathcal{P}_{i^{*}}}  > p^{(\text{opt})}_{\mathcal{P}_{i^{*}}} - (c + 1) \Delta_{\pi,\mathcal{P}_{i^{*}}}$.
But, we know that $p^{\text{(pert)}}_{\mathcal{P}_{i^{*}}}\leq  p^{\text{(opt)}}$, which leads to $ p^{\text{(opt)}} > p^{(\text{opt})}_{\mathcal{P}_{i^{*}}} - (c + 1) \Delta_{\pi,\mathcal{P}_{i^{*}}}$. Thus, we have $p^{(\text{opt})}_{\mathcal{P}_{i^{*}}} < p^{\text{(opt)}} + (c + 1) \Delta_{\pi,\mathcal{P}_{i^{*}}}$.  $\blacksquare$
\section{Proof of Theorem \ref{thm:mainresult1}} \label{app:mainresult1}
We consider the following instantaneous drift-plus-penalty expression denoted by $\mathcal{P}_{\tau,V} \triangleq \Delta(\tau+D) + V p_0(\tau)$
\begin{eqnarray}
\mathbb{E}\left[\mathcal{P}_{\tau,V}  \right] &=& \mathbb{E}\left[\mathcal{P}_{\tau,V}\left \vert \right.  \mathcal{E}_{\delta,\tau}^c \right] \Pr\{\mathcal{E}_{\delta,\tau}^c\} + \mathbb{E}\left[\mathcal{P}_{\tau,V}\left \vert \right. \mathcal{E}_{\delta,\tau} \right] \Pr\{\mathcal{E}_{\delta,\tau}\} \nonumber \\
&\leq& \mathbb{E}\left[\mathcal{P}_{\tau,V}\left \vert \right.  \mathcal{E}_{\delta,\tau}^c \right] + \mathbb{E}\left[\mathcal{P}_{\tau,V}\left \vert \right. \mathcal{E}_{\delta,\tau} \right] \Pr\{\mathcal{E}_{\delta,\tau}\}. \label{eq:dpp_bound}
\end{eqnarray}
where $\mathcal{E}_{\delta,\tau}$ is the error in slot $\tau \in \mathbb{N}$ of step $1$ of the ADPP \textbf{Algorithm} due to incorrectly detecting the ``right distribution," $\mathcal{P}_{i^*} \in \mathcal{P}_c$ (see \eqref{eq:detect_covering_alg}). 
Next, we will compute an upper bound on the second term in \eqref{eq:dpp_bound}, i.e., $\mathbb{E}\left[\mathcal{P}_{\tau,V}\left \vert \right. \mathcal{E}_{\delta,\tau} \right]$. Assume that the output of the \textbf{Algorithm} is $m^*$, and the corresponding induced probability be $\theta_m^* = 1$ if $m = m^*$, zero otherwise. Now, we consider the following drift-plus-penalty bound on the second term in  \eqref{eq:dpp_bound} conditioned on $\mathbf{Q(\tau)}$ 
\begin{eqnarray} 
\mathbb{E}\left[\mathcal{P}_{\tau,V}\left \vert \right. \mathcal{E}_{\delta,\tau}, \mathbf{Q(\tau)} \right] \stackrel{(a)}{\leq} H_\tau + V \sum_{m=1}^F \theta_m^* r_{0,\pi_\tau}^{(m)}    + \sum_{k=1}^K Q_k(\tau) \mathcal{C}_{k,\tau} 
\stackrel{(b)}{\leq} H_\tau + V p_{\text{max},0},
\end{eqnarray}
where $\mathcal{C}_{k,\tau} \triangleq \left[ \sum_{m=1}^F \theta_m^* r_{k,\pi_\tau}^{(m)} - c_k\right]$, $H_\tau\triangleq (1 + 2D) B_\tau$, $B_\tau$ is as defined in \eqref{eq:btau}, $(a)$ follows from Lemma $6$ of \cite{neely2016distributed}, and $(b)$ follows from the fact that there exists a strategy $m^{'}$ such that $\sum_{m=1}^F \theta_m^{'} r_{k,\pi_\tau}^{(m)} - c_k \leq 0$, and $p_{\text{max},0}$ is the maximum cost. Taking the expectation of the above with respect to $\mathbf{Q(\tau)}$ conditioned on $\mathcal{E}_{\delta, \tau}$ leads to 
$\textbf{Result I:  } \mathbb{E}\left[\mathcal{P}_{\tau,V}\left \vert \right. \mathcal{E}_{\delta,\tau}\right] \leq H_\tau + V p_{\text{max},0}.$
Applying Lemma $5$ of \cite{neely2016distributed} to the first term in \eqref{eq:dpp_bound} conditioned on $\mathbf{Q(\tau)}$, we get 
\begin{eqnarray} 
&&\hspace{-1.0cm}\mathbb{E} \left[\mathcal{P}_{\tau,V} \left \vert \right. \mathbf{Q}(\tau), \mathcal{E}_{\delta,\tau}^c \right] \leq H_\tau + V \sum_{m=1}^F \theta_m^* r_{0,\pi_\tau}^{(m)}  + \sum_{k=1}^K Q_k(\tau) \mathcal{C}_{k,\tau}\nonumber\\
&\hspace{-0.3cm}\leq&  \hspace{-0.4cm}H_\tau + V \sum_{m=1}^F \theta_m^* r_{0,\mathcal{P}_{i^*}}^{(m)} + V \sum_{m=1}^F \theta_m^* \sum_{\omega \in \Omega} \Delta^{(\omega)}_{\pi,\mathcal{P}_{i^*}} p_0(\mathbf{S}^{(m)}(\omega), \omega) \nonumber \\ && \hspace{-0.9cm}+ V \sum_{m=1}^F \theta_m^* \sum_{\omega \in \Omega} \Delta^{(\omega)}_{\pi_\tau,\pi} p_0(\mathbf{S}^{(m)}(\omega), \omega) + \sum_{k=1}^K Q_k(\tau) \mathcal{C}_{k,\tau}\nonumber \\
&\leq& \hspace{-0.1cm}H_\tau + V R_{0,\mathcal{P}_{i^*}} + V  {J}^\pi_{\pi_\tau}  + \sum_{k=1}^K Q_k(\tau) \left[ R_k^{*}(\tau) - c_k\right], \label{eq:dpp_bound_appendix}
\end{eqnarray}
where $\mathcal{C}_{k,\tau}$ is as defined earlier, $R_{0,\mathcal{P}_{i^*}} \triangleq\sum_{m=1}^F \theta_m^* r_{0,\mathcal{P}_{i^*}}^{(m)}$, $R_k^{*}(\tau) \triangleq \sum_{m=1}^F \theta_m^* r_{k,\pi_\tau}^{(m)}$, $\Delta^{(\omega)}_{\pi_\tau,\pi} \triangleq \abs{\pi_\tau(\omega) - \pi(\omega)}$, $H_\tau \triangleq B_\tau(1 + 2D)$, and ${J}^\pi_{\pi_\tau} \triangleq \max_{0\leq k \leq K} p_{\text{max},k} \left(\norm{\pi_\tau - \pi}_1 + \norm{\mathcal{P}_{i^*} - \pi}_1\right)$. Consider the following
\begin{eqnarray} 
\sum_{k=1}^K Q_k(\tau) \left[ R_k^{*}(\tau) - c_k\right]  \hspace{-0.3cm}
&\leq&\sum_{k=1}^K Q_k(\tau) \left[\sum_{m=1}^F \theta_m^* r_{k,\mathcal{P}_{i^*}}^{(m)} - c_k^{'} \right], \label{eq:temp1}
\end{eqnarray}
where $c_k^{'} \triangleq c_k - {J}^\pi_{\pi_\tau}$. We need $c_k > {J}^\pi_{\pi_\tau}$. The above inequality is obtained in a similar fashion to that of the first three terms in \eqref{eq:dpp_bound_appendix}, and using the fact that $\norm{\mathcal{P}_{i^*} - \pi}_1 < \delta$. Substituting \eqref{eq:temp1} in \eqref{eq:dpp_bound_appendix}, we get 
\begin{eqnarray} 
\hspace{-0.6cm}\mathbb{E} \left[\mathcal{P}_{\tau,V} \left \vert \right. \mathbf{Q}(\tau), \mathcal{E}_{\delta,\tau}^c\right] &\hspace{-0.4cm}\leq\hspace{-0.3cm}& H_\tau + V \sum_{m=1}^F \theta_m^* r_{0,\mathcal{P}_{i^*}}^{(m)} +  \sum_{k=1}^K Q_k(\tau)  \left[\sum_{m=1}^F \theta_m^* r_{k,\mathcal{P}_{i^*}}^{(m)} - c_k^{'} \right]. \label{eq:dpp_alg}
\end{eqnarray}
Note that the \textbf{Algorithm} chooses to minimize the right hand side of the above term when there is no error. Thus, choosing an alternative algorithm say $\theta_m$ will maximize the right hand side of \eqref{eq:dpp_alg}. Towards bounding the above further, let us choose a $\theta_m$ denoted $\theta_{m,\texttt{opt}}^{'}$ that optimally solves the problem $\mathbf{LP}_{\mathcal{P}_{i^*}}$ but with $c_k$ replaced by $c_k^{'}$. Further, let the corresponding optimal cost be $p^{'}_{\texttt{opt}}$. From \textbf{Assumption 2}, it follows that $p^{'}_{\texttt{opt}} < p^{(\text{opt})}_{\mathcal{P}_{i^{*}}} + c{J}^\pi_{\pi_\tau}$. Using the optimal  $\theta_{m,\texttt{opt}}^{'}$ in \eqref{eq:dpp_alg}, we get 
\begin{eqnarray}
 \mathbb{E} \left[\mathcal{P}_{\tau,V} \left \vert \right. \mathbf{Q}(\tau), \mathcal{E}_{\delta,\tau}^c\right] \stackrel{(a)}{\leq} V p^{'}_{\texttt{opt}} + H_\tau + V  {J}^\pi_{\pi_\tau} 
< V p^{(\text{opt})}_{\mathcal{P}_{i^{*}}} + V(c+1){J}^\pi_{\pi_\tau} + H_\tau,
\end{eqnarray}
where the inequality $(a)$ is obtained by noting that for $ \theta_m = \theta_{m,\texttt{opt}}^{'}$, $\left[\sum_{m=1}^F \theta_{m,\texttt{opt}}^{'} r_{k,\mathcal{P}_{i^*}}^{(m)} - c_k^{'} \right] < 0$, and $ p^{'}_{\texttt{opt}} = \sum_{m=1}^F \theta_{m,\texttt{opt}}^{'} r_{0,\mathcal{P}_{i^*}}^{(m)}$. Using $p^{(\text{opt})}_{\mathcal{P}_{i^{*}}} < p^{\text{(opt)}} + (c + 1) \Delta_{\pi,\mathcal{P}_{i^{*}}}$ from Theorem \ref{thm:lppit_popt_relation}, we get 
\begin{eqnarray} \label{eq:dpp_firstexpression}
{ \mathbb{E} \left[\mathcal{P}_{\tau,V} \left \vert \right. \mathbf{Q}(\tau), \mathcal{E}_{\delta,\tau}^c\right]}
\leq  V \mathcal{\psi_{\texttt{const}}} + {V(c+1){J}^\pi_{\pi_\tau} + H_\tau } ,
\end{eqnarray}
where $\mathcal{\psi_{\texttt{const}}}\triangleq p^{\text{(opt)}} + (c + 1) \Delta_{\pi,\mathcal{P}_{i^{*}}}$, and $\Delta_{\pi,\mathcal{P}_{i^{*}}}$ is as defined in Theorem \ref{thm:lppit_popt_relation}. Now, taking the expectation with respect to $\mathbf{Q(t)}$ conditioned on $\mathcal{E}^c_{\delta,\tau}$, we get
\begin{equation}
\textbf{Result II:  } { \mathbb{E} \left[\mathcal{P}_{\tau,V} \left \vert \right. \mathcal{E}_{\delta,\tau}^c\right]}
\leq  V \mathcal{\psi_{\texttt{const}}} + {V(c+1){J}^\pi_{\pi_\tau} + H_\tau }.
\end{equation}
Next, in the following, we compute an upper bound on the probability of error in  \eqref{eq:dpp_bound}, i.e., 
\begin{eqnarray}
\Pr\{\mathcal{E}_{\delta,\tau}\} &=& \Pr\left\{ \bigcup_{j: j \neq i^*} \frac{1}{w} \sum_{s=\tau-D - w + 1}^{\tau-D} \log \mathcal{P}_{j}(\omega(s)) > f_{\tau,D,w}\right\} \nonumber\\
&\leq& \sum_{j:j\neq i^*} \Pr\left\{ \frac{1}{w} \sum_{\tau=t-D}^{t-D-w} \log \left(\frac{\mathcal{P}_{j}(\omega(\tau))}{\mathcal{P}_{i^*}(\omega(\tau))}\right) > 0 \right\} \nonumber\\
&\stackrel{(a)}{\leq}& \sum_{j:j\neq i^*} \Pr\left\{ g_{\tau,D,w} - \mathcal{D}_{\tau,j} > -\mathcal{D}_{\tau,j} \right\},  
\end{eqnarray}
where $g_{\tau,D,w}\triangleq\frac{1}{w} \sum_{s=\tau-D}^{\tau-D-w} \log \left(\frac{\mathcal{P}_{j}(\omega(s))}{\mathcal{P}_{i^*}(\omega(s))}\right)$, $f_{\tau,D,w}\triangleq \frac{1}{w} \sum_{s=\tau-D}^{\tau-D-w}  \log \mathcal{P}_{i^*}(\omega(s))$, and $\mathcal{D}_{\tau,j}$ is as defined in the Theorem. In the above, $(a)$ follows from the union bound. By using the following boundedness property from \textbf{Assumption 3}, i.e.,  $$\log \left(\frac{\mathcal{P}_{j}(\omega(\tau))}{\mathcal{P}_{i^*}(\omega(\tau))}\right)  \leq \log\left(\frac{\alpha_\delta}{\beta_\delta}\right),$$ and using the Hoeffdings inequality, we get 
\begin{equation} 
\textbf{Result III:  } \Pr\{\mathcal{E}_{\delta,\tau}\} \leq  P^{(\tau)}_{e,\texttt{up}} \triangleq \exp\left\{- {2 \zeta_\delta \mathcal{D}_\tau^2 w} + \mathcal{H}(\mathcal{P},\delta) \right\},
\end{equation}
where $\zeta_\delta \triangleq {\left[\log\left(\frac{\alpha_\delta}{\beta_\delta}\right)\right]^2}$, $\mathcal{D}_\tau \triangleq \min_{j \neq i^*} \mathcal{D}_{\tau,j}$, and $\mathcal{H}(\mathcal{P},\delta)\triangleq\log M_\delta$ is the \emph{metric entropy}. Using \textbf{Result I}, \textbf{Result II} and \textbf{Result III} in \eqref{eq:dpp_bound}, we get 
\begin{equation} \nonumber
\mathbb{E}\left[\mathcal{P}_{\tau,V}  \right] \leq V \mathcal{\psi_{\texttt{const}}} + {V(c+1){J}^\pi_{\pi_\tau} + H_\tau } + \left(H_\tau + V p_{\text{max},0}\right) P^{(\tau)}_{e,\texttt{up}}. 
\end{equation}
Summing the above over all slots $\tau = 0,1,2,\ldots, t-1$, and dividing by $t$, we get 
\begin{eqnarray} \label{eq:dpp_sumtau_bound}
\frac{\mathbb{E}\left[ \mathcal{L}(t+D) - \mathcal{L}(D)\right]}{t} + \frac{V}{t} \sum_{\tau=0}^{t-1} \mathbb{E} p_0(\tau) &\leq& V \psi_{\texttt{const}} + V(c+1) \bar{J}_t  \nonumber\\
\hspace{-5cm}&& \hspace{-5cm}+~  \bar{H}_t + \frac{(1+2D)}{t} \sum_{\tau = 0}^{t-1} B_{\tau} P_{e,\texttt{up}}^{(\tau)} + \frac{V p_{\text{max},0}}{t}\sum_{\tau=0}^{t-1} P_{e,\texttt{up}}^{(\tau)}.
\end{eqnarray}
Using the fact that $\mathcal{L}(t+D) \geq 0$, and $\mathcal{L}(D) \leq C$ for some constant $C >0$, and after rearranging the terms, we get
\begin{eqnarray} \label{eq:diffpen_bound}
\mathbb{E} [\bar p_0(t)] - p^{\text{(opt)}}  &\leq& (c + 1) \Delta_{\pi,\mathcal{P}_{i^{*}}} + \psi_t(\delta),
\end{eqnarray}
where $\psi_t(\delta)$ is as defined in the theorem, and $\mathbb{E} [\bar p_0(t)] \triangleq \frac{1}{t}\sum_{\tau=0}^{t-1} \mathbb{E} p_0(\tau)$.
For any $\epsilon > 0$, choosing $\epsilon_0 = (c + 1) \Delta_{\pi,\mathcal{P}_{i^{*}}} + \psi_t(\delta) + \epsilon$ satisfies the bound on $\epsilon_0$ in Theorem \ref{thm:pacfirstresult}. Again from Theorem \ref{thm:pacfirstresult} and the bound in \eqref{eq:diffpen_bound}, we have $$\epsilon_{t,0}\triangleq \epsilon_0 + p^{(opt)} - \frac{1}{t}\sum_{\tau=0}^{t-1} \mathbb{E}p_0(\tau) \geq \epsilon.$$ Thus, using $\epsilon_{t,0} = \epsilon$ in \eqref{eq:mcdiarmid_pac1} and $v_t = t/u_t$, we get the following upper bound 
\begin{equation}
\Pr\left\{\frac{1}{t}\sum_{\tau=0}^{t-1} p_0(\tau) - p^{(opt)} >  \epsilon_0 \right\} \leq u_t \exp\left\{\frac{-2 \epsilon^2 t^2}{u_{\text{max},0}^2 u_t^2}\right\} + t \beta_{\texttt{ADPP},k}(u_t). 
\end{equation}
It is easy to verify that the above is less than or equal to $\gamma_0 > t \beta_{\texttt{ALG},k}(u_t)$ provided $t \in \mathcal{T}_{t,0}$, where $\mathcal{T}_{t,0}$ is as defined in the theorem. This proves \textbf{Part A} of the Theorem.

Multiplying \eqref{eq:dpp_sumtau_bound} by $t$, substituting for $\mathcal{L}(t+D)$, and using the fact that for all time slots $\tau$, there exists a constant $F$ such that $F  \geq  p^{\text{(opt)}} -  \mathbb{E} \left[ p_0(\tau) \right]$, we get (see \cite{neely2016distributed})
\begin{equation} \label{eq:queue_norm_bound}
\mathbb{E}\{\norm{\mathbf{Q}(t + D)}^2_2 \} \leq VFt + \Gamma_t,
\end{equation}
where $\Gamma_t$ is as defined in the theorem. From Jensen's Inequality, it follows from the above bound that
\begin{equation}
\frac{\mathbb{E}\{\abs{{Q_k}(t + D)} \}}{t} \leq Q_{\texttt{up}}(t) \triangleq \sqrt{\frac{VF}{t} + \frac{\Gamma_t}{ t^2}},
\end{equation}
for all $k=1,2,\ldots, K$. 
From Lemma $4$ of \cite{neely2016distributed}, we have $\mathbb{E}\{\bar{p}_k(t)\} \leq c_k + Q_{\texttt{up}}(t)$. 
Now, the right hand side of \eqref{eq:mcdiarmid_pac1} for $\epsilon_{t,k} = \epsilon$, $\epsilon_k = Q_{\texttt{up}}(t) + \epsilon$ is less than or equal to $\gamma_1$ provided $t \in \mathcal{T}_{t,1}$, where $\mathcal{T}_{t,1}$ is as defined in the theorem. 
 $\blacksquare$

\bibliographystyle{IEEEtran}
\bibliography{IEEEabrv,DPP2016}

\begin{thebibliography}{10}
\providecommand{\url}[1]{#1}
\csname url@samestyle\endcsname
\providecommand{\newblock}{\relax}
\providecommand{\bibinfo}[2]{#2}
\providecommand{\BIBentrySTDinterwordspacing}{\spaceskip=0pt\relax}
\providecommand{\BIBentryALTinterwordstretchfactor}{4}
\providecommand{\BIBentryALTinterwordspacing}{\spaceskip=\fontdimen2\font plus
\BIBentryALTinterwordstretchfactor\fontdimen3\font minus
  \fontdimen4\font\relax}
\providecommand{\BIBforeignlanguage}[2]{{%
\expandafter\ifx\csname l@#1\endcsname\relax
\typeout{** WARNING: IEEEtran.bst: No hyphenation pattern has been}%
\typeout{** loaded for the language `#1'. Using the pattern for}%
\typeout{** the default language instead.}%
\else
\language=\csname l@#1\endcsname
\fi
#2}}
\providecommand{\BIBdecl}{\relax}
\BIBdecl

\bibitem{ciftcioglu2013maximizing}
E.~N. Ciftcioglu, A.~Yener, and M.~J. Neely, ``Maximizing quality of
  information from multiple sensor devices: The exploration vs exploitation
  tradeoff,'' \emph{IEEE Journal of Selected Topics in Signal Processing},
  vol.~7, no.~5, pp. 883--894, 2013.

\bibitem{neely2013dynamic}
M.~J. Neely, ``Dynamic optimization and learning for renewal systems,''
  \emph{IEEE Trans. on Automatic Control}, vol.~58, no.~1, pp. 32--46, 2013.

\bibitem{neely2016distributed}
------, ``Distributed stochastic optimization via correlated scheduling,''
  \emph{IEEE/ACM Trans. on Networking}, vol.~24, no.~2, pp. 759--772, 2016.

\bibitem{urgaonkar2011optimal}
R.~Urgaonkar, B.~Urgaonkar, M.~J. Neely, and A.~Sivasubramaniam, ``Optimal
  power cost management using stored energy in data centers,'' in \emph{Proc.
  of the ACM SIGMETRICS joint Int. conf. on measurement and modeling of
  computer systems}.\hskip 1em plus 0.5em minus 0.4em\relax ACM, 2011, pp.
  221--232.

\bibitem{baghaie2010energy}
M.~Baghaie, S.~Moeller, and B.~Krishnamachari, ``Energy routing on the future
  grid: A stochastic network optimization approach,'' in \emph{IEEE Int. Conf.
  on Power System Technology (POWERCON)}, 2010, pp. 1--8.

\bibitem{neelyinfocom2015}
M.~J. Neely, ``Energy-aware wireless scheduling with near optimal backlog and
  convergence time tradeoffs,'' in \emph{IEEE Conference on Computer
  Communications (INFOCOM)}, April 2015, pp. 91--99.

\bibitem{neely2010stochastic}
------, ``Stochastic network optimization with application to communication and
  queueing systems,'' \emph{Synthesis Lectures on Communication Networks},
  vol.~3, no.~1, pp. 1--211, 2010.

\bibitem{neely2008fairness}
M.~J. Neely, E.~Modiano, and C.~P. Li, ``Fairness and optimal stochastic
  control for heterogeneous networks,'' \emph{IEEE/ACM Trans. On Networking},
  vol.~16, no.~2, pp. 396--409, 2008.

\bibitem{georgiadis2006resource}
L.~Georgiadis, M.~J. Neely, and L.~Tassiulas, \emph{Resource allocation and
  cross-layer control in wireless networks}.\hskip 1em plus 0.5em minus
  0.4em\relax Now Publishers Inc, 2006.

\bibitem{neely2006energy}
M.~J. Neely, ``Energy optimal control for time-varying wireless networks,''
  \emph{IEEE Trans. on Information Theory}, vol.~52, no.~7, pp. 2915--2934,
  2006.

\bibitem{neely2010efficient}
M.~J. Neely, A.~S. Tehrani, and A.~G. Dimakis, ``Efficient algorithms for
  renewable energy allocation to delay tolerant consumers,'' in \emph{First
  IEEE Int. Conf. on Smart Grid Communications (SmartGridComm),}, 2010, pp.
  549--554.

\bibitem{neely2010dynamic}
M.~J. Neely and L.~Huang, ``Dynamic product assembly and inventory control for
  maximum profit,'' in \emph{49th IEEE Conference on Decision and Control
  (CDC)}, 2010, pp. 2805--2812.

\bibitem{tassiulas1992stability}
L.~Tassiulas and A.~Ephremides, ``Stability properties of constrained queueing
  systems and scheduling policies for maximum throughput in multihop radio
  networks,'' \emph{IEEE Trans. on Automatic Control}, vol.~37, no.~12, pp.
  1936--1948, 1992.

\bibitem{tassiulas1993dynamic}
------, ``Dynamic server allocation to parallel queues with randomly varying
  connectivity,'' \emph{IEEE Trans. on Information Theory}, vol.~39, no.~2, pp.
  466--478, 1993.

\bibitem{han2014distributed}
Y.~Han, Y.~Zhu, and J.~Yu, ``A distributed utility-maximizing algorithm for
  data collection in mobile crowd sensing,'' \emph{Proc. IEEE GLOBECOM, Austin,
  USA}, 2014.

\bibitem{zhang_wcnc2013}
X.~Zhang, S.~Zhou, Z.~Niu, and X.~Lin, ``An energy-efficient user scheduling
  scheme for multiuser mimo systems with rf chain sleeping,'' in \emph{IEEE
  Wireless Communications and Networking Conference (WCNC)}, April 2013, pp.
  169--174.

\bibitem{wei2015sample}
X.~Wei, H.~Yu, and M.~J. Neely, ``A sample path convergence time analysis of
  drift-plus-penalty for stochastic optimization,'' \emph{arXiv preprint
  arXiv:1510.02973}, 2015.

\bibitem{neely_maxweight_unknownenvironment_2012}
M.~J. Neely, S.~T. Rager, and T.~F.~L. Porta, ``Max weight learning algorithms
  for scheduling in unknown environments,'' \emph{IEEE Trans. on Automatic
  Control}, vol.~57, no.~5, pp. 1179--1191, May 2012.

\bibitem{van2000applications}
S.~A. Van~de Geer, \emph{Applications of empirical process theory}.\hskip 1em
  plus 0.5em minus 0.4em\relax Cambridge University Press Cambridge, 2000,
  vol.~91.

\bibitem{csiszar2011information}
I.~Csiszar and J.~K{\"o}rner, \emph{Information theory: coding theorems for
  discrete memoryless systems}.\hskip 1em plus 0.5em minus 0.4em\relax
  Cambridge University Press, 2011.

\bibitem{cover2012elements}
T.~M. Cover and J.~A. Thomas, \emph{Elements of information theory}.\hskip 1em
  plus 0.5em minus 0.4em\relax John Wiley \& Sons, 2012.

\bibitem{kuznetsov2014generalization}
V.~Kuznetsov and M.~Mohri, ``Generalization bounds for time series prediction
  with non-stationary processes,'' in \emph{Int. Conf. on Algorithmic Learning
  Theory}.\hskip 1em plus 0.5em minus 0.4em\relax Springer, 2014, pp. 260--274.

\end{thebibliography}

\end{document}